\documentclass[conference]{IEEEtran}
\usepackage{cite}
\usepackage{amsmath,amssymb,amsfonts}
\usepackage{bm}
\usepackage{algorithmic}
\usepackage{graphicx}
\usepackage{textcomp}
\usepackage{booktabs}
\usepackage{multirow}
\usepackage{xcolor}
\usepackage[ruled, linesnumbered]{algorithm2e}
\usepackage{threeparttable}
\usepackage[hmargin=0.75in, vmargin=0.75in]{geometry}

\IEEEoverridecommandlockouts
\graphicspath{{figures/}}
\def\BibTeX{{\rm B\kern-.05em{\sc i\kern-.025em b}\kern-.08em
    T\kern-.1667em\lower.7ex\hbox{E}\kern-.125emX}}

\begin{document}
\newgeometry{hmargin=0.75in, vmargin=1in}

\title{Decentralized Time and Energy-Optimal Control of Connected and
Automated Vehicles in a Roundabout\\
\thanks{%
This work was supported in part by NSF under grants ECCS-1931600,
DMS-1664644 and CNS-1645681, by ARPAE under grant DE-AR0001282, by AFOSR
under grant FA9550-19-1-0158, and by the MathWorks. The authors are with the
Division of Systems Engineering and Center for Information and Systems
Engineering, Boston University, Brookline, MA, 02446, USA \{xky, cgc,
xiaowei\}@bu.edu} }
\author{\IEEEauthorblockN{Kaiyuan Xu} \and 
\IEEEauthorblockN{Christos G.
Cassandras} \and \IEEEauthorblockN{Wei Xiao} }
\maketitle

\begin{abstract}
The paper considers the problem of controlling Connected and Automated
Vehicles (CAVs) traveling through a three-entry roundabout so as to jointly
minimize both the travel time and the energy consumption while providing
speed-dependent safety guarantees, as well as satisfying velocity and
acceleration constraints. We first design a systematic approach to
dynamically determine the safety constraints and derive the unconstrained
optimal control solution. A joint optimal control and barrier function
(OCBF) method is then applied to efficiently obtain a controller that
optimally track the unconstrained optimal solution while guaranteeing all
the constraints. Simulation experiments are performed to compare the optimal
controller to a baseline of human-driven vehicles showing effectiveness
under symmetric and asymmetric roundabout configurations, balanced and
imbalanced traffic rates and different sequencing rules for CAVs.
\end{abstract}


\section{Introduction}

The performance of traffic networks critically depends on the control of
conflict areas such as intersections, roundabouts and merging roadways which
are the main bottlenecks in these networks. The economic loss caused by
congestion in these areas has been well documented \cite{rios2016survey}.
Coordinating and controlling vehicles in these conflict areas is a
challenging problem in terms of safety, congestion, and energy consumption 
\cite{chen2015cooperative, tideman2007review}. The emergence of Connected
and Automated Vehicles (CAVs) provides a promising solution to this problem
through better information utilization and more precise trajectory design.
The automated control of vehicles has gained increasing attention with the
development of new traffic infrastructure technologies \cite{li2013survey}
and, more recently, CAVs \cite{rios2016survey}.

Both centralized and decentralized methods have been studied to deal with
the control and coordination of CAVs at conflict areas. Centralized
mechanisms are often used in forming platoons in merging problems \cite%
{xu2019grouping} and determining passing sequences at intersections \cite%
{xu2020bi}. These approaches tend to work better when the safety constraints
are independent of speed and they generally require significant computation
resources, especially when traffic is heavy. They are also not easily
amenable to disturbances.

Decentralized mechanisms restrict all computation to be done on board each
vehicle with information sharing limited to a small number of neighbor
vehicles \cite{milanes2010automated, rios2015online, bichiou2018developing,
hult2016coordination}. Optimal control problem formulations are used in some
of these approaches, with Model Predictive Control (MPC) techniques employed
as an alternative to account for additional constraints and to compensate
for disturbances by re-evaluating optimal actions \cite{cao2015cooperative,
mukai2017model, nor2018merging}. The objectives in such problem formulations
typically target the minimization of acceleration or the maximization of
passenger comfort (measured as the acceleration derivative or jerk). An
alternative to MPC has recently been proposed through the use of Control
Barrier Functions (CBFs) \cite{xiao2020bridging, xiao2019decentralized}
which provide provable guarantees that safety constraints are always
satisfied.

In this paper, we build on the use of optimal control and CBF-based methods
in unsignalized intersections \cite{zhang2019decentralized} and merging \cite%
{xiao2021decentralized} to study roundabouts with all traffic consisting of
CAVs. There are several similarities between merging, intersections and
roundabouts problems. The single-lane merging problem \cite%
{xiao2021decentralized} contains a single Merging Point (MP) where safety
constraints must be guaranteed, while CAVs follow the same moving direction
in each lane. In intersection problems, CAVs have a number of possible paths
which conflict at multiple MPs restricted to a small area. In a roundabout,
CAVs have the same moving direction (either clockwise or counterclockwise)
but multiple possible paths which cross at multiple MPs. A roundabout
problem can be dealt with as either a whole system like an intersection or
it can be decomposed into several coupled merging problems.

Roundabouts are important components of a traffic network because they
usually perform better than typical intersections in terms of efficiency and
safety \cite{flannery1997operational}. However, they can become sgnificant
bottleneck points as the traffic rate increases due to an inappropriate
priority system, resulting in significant delays when the circulating flow
is heavy. Previous studies mainly focus on conventional vehicles and try to
solve the problem through improved road design or traffic signal control 
\cite{martin2016benefits, yang2004new, xu2016multi}. More recently, however,
researchers have proposed methods for decentralized optimal control of CAVs
in a roundabout. The roundabout problem is formulated as \restoregeometry 
\noindent an optimal control 
problem with an analytical solution provided in \cite{zhao2018optimal}. The
problem is decomposed so that first the minimum travel time is solved under
the assumption that all vehicles use the same maximum speed within the
roundabout. Then, fixing this time, the control input that minimizes the
energy consumption is derived analytically. The general framework for
decentralized optimal control of CAVs used in urban intersections is
implemented in a roundabout problem in \cite{chalaki2020experimental}. The
analysis is similar to \cite{zhao2018optimal} except that there is no
circulating speed assumption.

In this paper, we formulate an optimal control problem for controlling CAVs
traveling through a roundabout. Unlike \cite{zhao2018optimal,
chalaki2020experimental}, we \emph{jointly} minimize both the travel time
and the energy consumption and also consider speed-dependent safety
constraints at a set of MPs rather than merging zones (which makes solutions
less conservative by improving roadway utilization). In addition, to improve
computational efficiency, we adopt the joint Optimal Control and Barrier
Function (OCBF) introduced in \cite{xiao2019decentralized}: we first derive
the optimal solution when no constraints become active and subsequently
optimally track this solution while also guaranteeing the satisfaction of
all constraints through the use of CBFs. We first assume a
First-In-First-Out (FIFO) sequencing policy over the entire system. We then
divide the roundabout into separate merging problems so as to introduce
different resequencing rules depending on the MP. We will show that the FIFO
policy does not perform well in many \textquotedblleft
asymmetric\textquotedblright\ configurations and explore an alternative
sequencing policy, termed Shortest Distance First (SDF), which our
experimental results show to be superior to FIFO. However, a systematic
study of the effect of the passing order is still the subject of ongoing
research. 

The paper is organized as follows. In Section II, the roundabout problem is
formulated as an optimal control problem with safety constraints. In Section
III, a decentralized framework is provided to determine the safety
constraints related to a given CAV. An OCBF controller is designed in
Section IV while simulation results are presented in Section V showing
significant improvements in the performance of the OCBF controller compared
to a baseline of human-driven vehicles. Finally, in Section VI we provide
conclusions and future research. 

\section{Problem Formulation}

\label{sec:problem} 
\begin{figure}[htb]
\centering
\includegraphics[width = 0.8\linewidth]{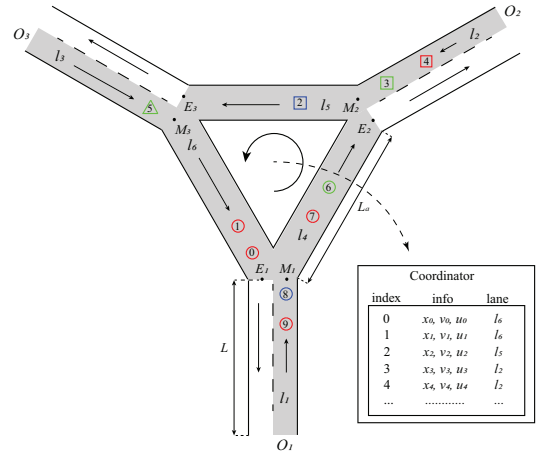}
\caption{A roundabout with 3 entries}
\label{fig:model}
\end{figure}

We initiate our study of roundabouts by considering a
single-lane triangle-shaped roundabout with 3 entries and 3 exits as shown
in Fig. \ref{fig:model}. We consider the case where all traffic consists of
CAVs which randomly enter the roundabout from three different origins $%
O_{1},O_{2}$ and $O_{3}$ and have randomly assigned exit points $E_{1},E_{2}$
and $E_{3}$. The gray road segments which include the triangle and three
entry roads form the Control Zone (CZ) where CAVs can share information and
thus be automatically controlled. We assume all CAVs move in a
counterclockwise way in the CZ. The entry road segments are connected with
the triangle at the three Merging Points (MPs) where CAVs from different
road segments may potentially collide with each other. The MPs are labeled
as $M_{1}$, $M_{2}$ and $M_{3}$. 
We assume that each road segment has one single lane (extensions to multiple
lanes and MPs are possible following the analysis in \cite%
{xiao2020decentralized}) The three entry road segments which are labeled as $%
l_{1}$, $l_{2}$ and $l_{3}$ have the same length $L$, while the road
segments in the triangle which are labeled as $l_{4}$, $l_{5}$ and $l_{6}$
have the same length $L_{a}$ (extensions to different lengths are
straightforward). In Fig. \ref{fig:model}, a circle, square and triangle
represent entering from $O_{1}$, $O_{2}$ and $O_{3}$ respectively. The color
red, green and blue represents exiting from $E_{1}$, $E_{2}$ and $E_{3}$
respectively. The full trajectory of a CAV in terms of the MPs it must go
through can be determined by its entry and exit points.

A coordinator, i.e., a Road Side Unit (RSU) associated with the roundabout,
maintains a passing sequence for all CAVs and records the information of
each CAV. The CAVs communicate with the coordinator but are not controlled
by it. All control inputs are evaluated on board each CAV in a \emph{%
decentralized} way. Each CAV is assigned a unique index upon arrival at the
CZ according to the passing order. The most common scheme for maintaining a
passing sequence is the First-In-First-Out (FIFO) policy according to each
CAV's arrival time at the CZ. The FIFO rule is one of the simplest schemes,
yet works well in many occasions as also shown in \cite{xu2020comparison}.
For simplicity, in what follows we use the FIFO queue, but we point out that
any passing order policy may be used, such as the Dynamic Resequencing (DR)
method in \cite{zhang2018decentralized}. We also introduce one such
alternative in Section \ref{sec:method}.

Let $S(t)$ be the set of CAV indices in the coordinator queue table at time $%
t$. The cardinality of $S(t)$ is denoted as $N(t)$. When a new CAV arrives,
it is allocated the index $N(t)+1$. Each time a CAV $i$ leaves the CZ, it is
dropped and all CAV indices larger than $i$ decrease by one. When CAV $i\in
S(t)$ is traveling in the roundabout, there are several important \emph{%
events} whose times are used in our analysis: $(i)$ CAV $i$ enters the CZ at
tme $t_{i}^{0} $, $(ii)$ CAV $i$ arrives at MP $M_{k}$ at time $t_{i}^{k},$ $%
k\in \{1,2,3\}$, $(iii)$ CAV $i$ leaves the CZ at time $t_{i}^{f}$. Based on
this setting, we can formulate an optimal control problem as described next.

\textbf{Vehicle Dynamics} Denote the distance from the origin $O_{j},j\in
\{1,2,3\}$ to the current location of CAV $i$ along its trajectory as $%
x_{i}^{j}(t)$. Since the CAV's unique identity $i$ contains the information
about the CAV's origin $O_{j}$, we can use $x_{i}(t)$ instead of $%
x_{i}^{j}(t)$ (without any loss of information) to describe the vehicle
dynamics as 
\begin{equation}
\left[ 
\begin{matrix}
\dot{x}_{i}(t) \\ 
\dot{v}_{i}(t)%
\end{matrix}%
\right] =\left[ 
\begin{matrix}
v_{i}(t) \\ 
u_{i}(t)%
\end{matrix}%
\right]   \label{equ:dynamics}
\end{equation}%
where $v_{i}$ is the velocity CAV $i$ along its trajectory and $u_{i}$ is
the acceleration (control input).

\textbf{Objective 1} Minimize the travel time $J_{i,1}=t_{i}^{f}-t_{i}^{0}$
where $t_{i}^{0}$ and $t_{i}^{f}$ are the times CAV $i$ enters and exits the
CZ.

\textbf{Objective 2} Minimize energy consumption: 
\begin{equation}
J_{i,2}=\int_{t_{i}^{0}}^{t_{i}^{f}}C_{i}(u_{i}(t))dt
\end{equation}%
where $C_{i}(\cdot )$ is a strictly increasing function of its argument.
Since the energy consumption rate is a monotonic function of the
acceleration, we adopt this general form to achieve energy efficiency.

\textbf{Constraint 1} (Rear-end safety constraint) Let $i_{p}$ denote the
index of the CAV which immediately precedes CAV $i$ on road segment $l_{k}$.
The distance between $i_{p}$ and $i$ $z_{i,i_{p}}(t)\equiv
x_{i_{p}}(t)-x_{i}(t)$ should be constrained by a speed-dependent
constraint: 
\begin{equation}
z_{i,i_{p}}(t)\geq \varphi v_{i}(t)+\delta ,\text{ \ }\forall t\in \lbrack
t_{i}^{0},t_{i}^{f}],\text{ \ }\forall i\in S(t)  \label{equ:constraint1}
\end{equation}%
where $\varphi $ denotes the reaction time (as a rule, $\varphi =1.8$ is
suggested, see \cite{vogel2003comparison}), $\delta $ denotes the minimum
safety distance (in general, we may use $\delta _{i}$ to make this distance
CAV-dependent but will use a fixed $\delta $ for simplicity). The index of
the preceding CAVs index $i_{p}$ may change due to road segment changing
events and is determined by the method described later in section \ref{sc}.

\textbf{Constraint 2} (Safe merging constraint) Let $t_{i}^{k},$ $k\in
\{1,2,3\}$ be the arrival time of CAV $i$ at MP $M_{k}$. Let $i_{m}$ denote
the index of the CAV that CAV $i$ may collide with when arriving at its next
MP $M_{k}$. The distance between $i_{m}$ and $i$ $z_{i,i_{m}}(t)\equiv
x_{i_{m}}(t)-x_{i}(t)$ is constrained by: 
\begin{equation}
z_{i,i_{m}}(t_{i}^{k})\geq \varphi v_{i}(t_{i}^{k})+\delta ,\text{ \ }%
\forall i\in S(t),\text{ \ }k\in \{1,2,3\}  \label{equ:constraint2}
\end{equation}%
where $i_{m}$ can be determined and updated by the method described in
section \ref{sc}.

\textbf{Constraint 3} (Vehicle limitations) The CAVs are also subject to
velocity and acceleration constraints due to physical limitations or road
rules: 
\begin{equation}
\begin{split}
v_{i,\min }& \leq v_{i}(t)\leq v_{i,\max },\forall t\in \lbrack
t_{i}^{0},t_{i}^{f}],\forall i\in S(t) \\
u_{i,\min }& \leq u_{i}(t)\leq u_{i,\max },\forall t\in \lbrack
t_{i}^{0},t_{i}^{f}],\forall i\in S(t)
\end{split}
\label{equ:constraint3}
\end{equation}%
where $v_{i,\max }>0$ and $v_{i,\min }\geq 0$ denote the maximum and minimum
speed for CAV $i$, $u_{i,\max }<0$ and $u_{i,\min }<0$ denote the maximum
and minimum acceleration for CAV $i$. We further assume common speed limits
dictated by the traffic rules at the roundabout, i.e. $v_{i,\min }=v_{\min }$%
, $v_{i,\max }=v_{\max }$.

Similar to previous work \cite{xiao2021decentralized}, we construct a convex
combination of the two objectives above: 
\begin{equation}
J_{i}=\int_{t_{i}^{f}}^{t_{i}^{0}}\left[ \alpha +(1-\alpha )\frac{\frac{1}{2}%
u_{i}^{2}(t)}{\frac{1}{2}\max \{u_{\max {}}^{2},u_{\min {}}^{2}\}}\right] dt
\end{equation}%
where $J_{i,1}$ and $J_{i,2}$ are combined with $\alpha \in \lbrack 0,1]$
after proper normalization. Here, we simply choose the quadratic function $%
C_{i}(u_{i})=\frac{1}{2}u_{i}^{2}(t)$. If $\alpha =1$, the problem
degenerates into a minimum traveling time problem. If $\alpha =0$, it
degenerates into a minimum energy consumption problem.

By defining $\beta \equiv \frac{\alpha }{2(1-\alpha )}\max \{u_{\max
{}}^{2},u_{\min {}}^{2}\},\alpha \in \lbrack 0,1)$ and proper scaling, we
can rewrite this minimization problem as 
\begin{equation}
J_{i}(u_{i})=\beta (t_{i}^{f}-t_{i}^{0})+\int_{t_{i}^{0}}^{t_{i}^{f}}\frac{1%
}{2}u_{i}^{2}(t)dt  \label{equ:obj}
\end{equation}%
where $\beta $ is the weight factor derived from $\alpha $. Then, we can
formulate the optimal control problem as follows:

\textbf{Problem 1}: For each CAV $i$ following the dynamics %
\eqref{equ:dynamics}, find the optimal control input $u_{i}(t)$ that
minimizes \eqref{equ:obj} subject to constraints \eqref{equ:dynamics}, %
\eqref{equ:constraint1}, \eqref{equ:constraint2}, \eqref{equ:constraint3},
the initial condition $x_{i}(t_{i}^{0})=0$, and given $t_{i}^{0}$, $%
v_{i}^{0} $ and $x_{i}(t_{i}^{f})$.

\section{Decentralized Control Framework}

\label{sec:method}

Compared to the single-lane merging or intersection control problems where
the constraints are determined and fixed immediately when CAV $i$ enters the
CZ, the main difficulty in a roundabout is that the constraints generally
change after every event (defined earlier). In particular, for each CAV $i$
at time $t$ only the merging constraint related to the next MP ahead is
considered. In other words, we need to determine at most one $i_{p}$ to
enforce \eqref{equ:constraint1} and one $i_{m}$ to enforce %
\eqref{equ:constraint2} at any time instant. There are two different ways to
deal with this problem: $(i)$ Treat the system as a single CZ with three MPs
with knowledge of each CAV's sequence of MPs, or $(ii)$ Decompose the
roundabout into three separate merging problems corresponding to the three
MPs, each with a separate CZ. The first approach is more complex and heavily
relies on the CAV sequencing rule used. If FIFO is applied, it is likely to
perform poorly in a large roundabout, because a new CAV may experience a
large delay in order to preserve the FIFO passing sequence. The second
approach is easier to implement but may cause congestion in a small
roundabout due to the lack of space for effective control at each separate
CZ associated with each MP. 

In order to solve \textbf{Problem 1} for each CAV $i$, we need to first
determine the corresponding $i_{p}$ and $i_{m}$ (when they exist) required
in the safety constraints (\ref{equ:constraint1}) and (\ref{equ:constraint2}%
). Once this task is complete and (\ref{equ:constraint1}) and (\ref%
{equ:constraint2}) are fully specified, then \textbf{Problem 1} can be
solved. In what follows, this first task is accomplished through a method
designed to determine the constraints in an event-driven manner which can be
used in either of the two approaches above and for any desired sequencing
policy. An extended queue table, an example of which is shown in Table. \ref%
{tab:ecq} corresponding to Fig. \ref{fig:model}, is used to record the
essential state information and identify all conflicting CAVs. We specify
the state-updating mechanism for this queue table so as to determine for
each CAV $i$ the corresponding $i_{p}$ and $i_{m}$. Then, we focus on the
second approach and in Section IV develop a general algorithm for solving 
\textbf{Problem 1} based on the OCBF method.

\subsection{The Extended Coordinator Queue Table}

Starting with the coordinator queue table shown in Fig. \ref{fig:model}, we
extend it to include 6 additional columns for each CAV $i$. The precise
definitions are given below: 
\begin{table}[tbh]
\caption{The extended coordinator queue table $S(t)$}
\label{tab:ecq}\centering
\begin{tabular}{|c|c|c|c|c|c|c|c|c|}
\hline
\multicolumn{9}{|l|}{$S(t)$} \\ \hline
idx & state & curr. & ori. & 1st MP & 2nd MP & 3rd MP & $i_p$ & $i_m$ \\ \hline
0 & $\bm x_0$ & $l_6$ & $l_1$ & $M_1$, M & $M_2$, M & $M_3$, M &  &  \\ 
\hline
1 & $\bm x_1$ & $l_6$ & $l_1$ & $M_1$, M & $M_2$, M & $M_3$, M & 0 &  \\ 
\hline
2 & $\bm x_2$ & $l_5$ & $l_2$ & $M_2$, M &  &  &  &  \\ \hline
3 & $\bm x_3$ & $l_2$ & $l_2$ & $M_2$ & $M_3$ & $M_1$ &  & 2 \\ \hline
4 & $\bm x_4$ & $l_2$ & $l_2$ & $M_2$ & $M_3$ &  & 3 &  \\ \hline
5 & $\bm x_5$ & $l_3$ & $l_3$ & $M_3$ & $M_1$ &  &  & 1 \\ \hline
6 & $\bm x_6$ & $l_4$ & $l_1$ & $M_1$, M &  &  &  &  \\ \hline
7 & $\bm x_7$ & $l_4$ & $l_1$ & $M_1$, M & $M_2$ & $M_3$ & 6 & 4 \\ \hline
8 & $\bm x_8$ & $l_1$ & $l_1$ & $M_1$ & $M_2$ &  &  & 7 \\ \hline
9 & $\bm x_9$ & $l_1$ & $l_1$ & $M_1$ & $M_2$ & $M_3$ & 8 &  \\ \hline
\end{tabular}%
\end{table}

\begin{itemize}
\item \emph{idx}: Unique CAV index, which allows us to determine the order
in which the CAV will leave the roundabout according to some policy (e.g.,
FIFO in Table I).

\item \emph{state}: CAV state $\bm x_{i}=(x_{i},v_{i})$ where $x_{i}$
denotes the distance from the entry point to the location of CAV $i$ along
its current road segment.

\item \emph{curr.}: Current CAV road segment, which allows us to determine
the rear-end safety constraints.

\item \emph{ori.}: Original CAV road segment, which allows us to determine
its relative position when in road segment \emph{curr}. 

\item \emph{1st-3rd MP}: These columns record all the MPs on the
CAV trajectory. If a CAV has already passed an MP, this MP is marked with an 
$M$. Otherwise, it is unmarked. As a CAV may not pass all three MPs in the
roundabout, some of these columns may be blank.

\item \emph{$i_{p}$}: Index of the CAV that immediately precedes CAV $i$ in
the same road segment (if such a CAV exists).

\item \emph{$i_{m}$}: Index of the CAV that may conflict with CAV $i$ at the
next MP. CAV $i_{m}$ is the last CAV that passes the MP ahead of CAV $i$.
Note that if $i_{m}$ and $i$ are in the same road segment, then $i_{m}$ $%
(=i_{p})$ is the immediately preceding CAV. In this case, the safe merging
constraint is redundant and need not be included.
\end{itemize}

\textbf{Event-driven Update Process for $S(t)$}: The extended coordinator
queue table $S(t)$ is updated whenever an event (as defined earlier) occurs.
Thus, there are three different update processes corresponding to each
triggering event:

\begin{itemize}
\item A new CAV enters the CZ: The CAV is indexed and added to the bottom of
the queue table.

\item CAV $i$ exits the CZ: All information of CAV $i$ is removed. All rows
with index larger than $i$ decrease their index values by 1.

\item CAV $i$ passes an MP: Mark the MP with $M$ and update the current road
segment value \emph{curr} of CAV $i$ with the one it is entering.
\end{itemize}

\subsection{Determination of Safety Constraints}

\label{sc} Recall that for each CAV $i$ in the CZ, we need to consider two
different safety constraints (\ref{equ:constraint1}) and (\ref%
{equ:constraint2}). First, by looking at each row $j<i$ and the
corresponding current road segment value \emph{curr}, CAV $i$ can determine
its immediately preceding CAV $i_{p}$ if one exists. This fully specifies
the rear-end safety constraint (\ref{equ:constraint1}). Next, we determine
the CAV which possibly conflicts with CAV $i$ at the next MP it will pass so
as to specify the safe merging constraint (\ref{equ:constraint2}). To do so,
we find in the extended queue table the last CAV $j<i$ that will pass or has
passed the same MP as CAV $i$. In addition, if the CAV is in the same road
segment as CAV $i$, it coincides with the preceding CAV $i_{p}$. Otherwise,
we find $i_{m}$, if it exists. As an example, in Table \ref{tab:ecq} (a
snapshot of Fig. \ref{fig:model}), CAV 8 has no immediate preceding CAV in $%
l_{1}$, but it needs to yield to CAV 7 (although CAV 7 has already passed $%
M_{3}$, when CAV 8 arrives at $M_{3}$ there needs to be adequate space
between CAV 7 and 8 for CAV 8 to enter $l_{4}$). CAV 9 however, only needs
to satisfy its rear-end safety constraint with CAV 8.

It is now clear that we can use the information in $S(t)$ in a systematic
way to determine both $i_{p}$ in \eqref{equ:constraint1} and $i_{m}$ in %
\eqref{equ:constraint2}. Thus, there are two functions $i_{p}(e)$ and $%
i_{m}(e)$ which need to be updated after event $e$ if this event affects CAV 
$i$. The index $i_{p}$ can be easily determined by looking at rows $j<i$ in
the extended queue table until the first one is found with the same value 
\emph{curr} as CAV $i$. For example, CAV 9 searches for its $i_{p}$ from CAV
8 to the top and sets $i_{p}=8$ as CAV 8 has the \emph{curr} value $l_{1}$.
Next, the index $i_{m}$ is determined. To do this, CAV $i$ compares its MP
information to that of each CAV in rows $j<i$. The process terminates the
first time that any one of the following two conditions is satified:

\begin{itemize}
\item The MP information of CAV $i_{m}$ \emph{matches} CAV $i$. We define $%
i_{m}$ to \textquotedblleft match\textquotedblright\ $i$ if and only if the
last marked MP or the first unmarked MP of CAV $i_{m}$ is the same as the
first unmarked MP of CAV $i$.

\item All prior rows $j<i$ have been looked up and none of them matches the
MP information of CAV $i$.
\end{itemize}

Combining the two updating processes for $i_{p}$ and $i_{m}$ together, there
are four different cases as follows:

\textbf{1. Both $i_{p}$ and $i_{m}$ exist.} In this case, there are two
possibilities: $(i)$ $i_{p}\neq i_{m}$. CAV $i$ has to satisfy the safe
merging constraint \eqref{equ:constraint2} with $i_{p}<i$ and also satisfy
the rear-end safety constraint \eqref{equ:constraint1} with $i_{m}<i$. For
example, for $i=7$, we have $i_{p}=6$ and $i_{m}=4$ ($M_{2}$ is the first
unmarked MP for CAV 7 and that matches the first unmarked MP for CAV 4). $%
(ii)$ $i_{p}=i_{m}$. CAV $i$ only has to follow $i_{p}$ and satisfy the
rear-end safety constraint \eqref{equ:constraint1} with respect to $i_{p}$.
Thus, there is no safe merging constraint for CAV $i$ to satisfy. For
example, $i=4$ and $i_{p}=i_{m}=3$.

\textbf{2. Only $i_{p}$ exists.} In this case, there is no safe merging
constraint for CAV $i$ to satisfy. CAV $i$ only needs to follow the
preceding CAV $i_{p}$ and satisfy the rear-end safety constraint %
\eqref{equ:constraint1} with respect to $i_{p}$. For example, $i=1$ and $%
i_{p}=0$.

\textbf{3. Only $i_{m}$ exists.} In this case, CAV $i$ has to satisfy the
safe merging constraint (\ref{equ:constraint2}) with the CAV $i_{m}$ in $S(t)
$. There is no preceding CAV $i_{p}$, thus there is no rear-end safety
constraint. For example, $i=5$, $i_{m}=1$ ($M_{3}$ is the first unmarked MP
for CAV 5 and that matches the last marked MP for CAV 1 with no other match
for $j=4,3,2$).

\textbf{4. Neither $i_{p}$ nor $i_{m}$ exists.} In this case, CAV $i$ does
not have to consider any safety constraints. For example, $i=2$.

\subsection{Sequencing Policies with Sub-coordinator Queue Tables}

Thus far, we have assumed a FIFO sequencing policy in the whole roundabout
and defined a systematic process for updating $i_{p}(e)$ and $i_{m}(e)$
after each event $e$, hence, the entire extended queue table $S(t)$.
However, FIFO may not be a good sequencing policy if applied to the whole
roundabout. In order to allow possible resequencing when a CAV passes an MP,
we introduce next a sub-coordinator queue table $S_{k}(t)$ associated with
each $M_{k}$, $k=1,2,3$. $S_{k}(t)$ coordinates all the CAVs for which $M_{k}
$ is the next MP to pass or it is the last MP that they have passed. We
define CZ$_{k}$ as the CZ corresponding to $M_{k}$ that consists of the
three road segments directly connected to $M_{k}$. A sub-coordinator queue
table can be viewed as a subset of the extended coordinator queue table
except that the CAVs are in different order in the two tables. As an
example, Table \ref{tab:scq1} (a snapshot of Fig. \ref{fig:model}) is the
sub-coordinator queue table corresponding to $M_{1}$ (in this case, still
based on the FIFO policy).

\begin{table}[htb]
\caption{The sub-coordinator queue table $S_1(t)$}
\label{tab:scq1}\centering
\begin{tabular}{|c|c|c|c|c|c|c|c|c|}
\hline
\multicolumn{9}{|l|}{$S_1(t)$} \\ \hline
idx & info & curr. & ori. & 1st MP & 2nd MP & 3rd MP & $i_p$ & $i_m$ \\ 
\hline
6 & $\bm x_6$ & $l_4$ & $l_1$ & $M_1$, M &  &  &  &  \\ \hline
7 & $\bm x_7$ & $l_4$ & $l_1$ & $M_1$, M & $M_2$ & $M_3$ & 6 &  \\ \hline
8 & $\bm x_8$ & $l_1$ & $l_1$ & $M_1$ & $M_2$ &  &  & 7 \\ \hline
0 & $\bm x_0$ & $l_6$ & $l_1$ & $M_1$, M & $M_2$, M & $M_3$, M &  &  \\ 
\hline
1 & $\bm x_1$ & $l_6$ & $l_1$ & $M_1$, M & $M_2$, M & $M_3$, M & 0 &  \\ 
\hline
9 & $\bm x_9$ & $l_1$ & $l_1$ & $M_1$ & $M_2$ & $M_3$ & 8 &  \\ \hline
\end{tabular}%
\end{table}

\textbf{Event-driven Update Process for $S_{k}(t)$}: The sub-coordinator
queue table $S_{k}(t)$ is updated as follows after each event that has
caused an update of the extended coordinator queue table $S(t)$ as follows:

\begin{itemize}
\item For each CAV $i$ in a sub-coordinator queue table $S_{k}(t)$, update
its information depending on the event observed: $(i)$ A new CAV $j$ enters
CZ$_{k}$ (either from an enrty point to the roundabout CZ or a MP passing
event): Add a new row to $S_{k}(t)$ and resequence the sub-coordinator queue
table according to the sequencing policy used. $(ii)$ CAV $j$ exits CZ$_{k}$%
: Remove all the information of CAV $j$ from $S_{k}(t)$.%

\item Determine $i_{p}$ and $i_{m}$ in each sub-coordinator queue table with
the same method as described in section \ref{sc}.

\item Update CAV $j$'s $i_{p}$ and $i_{m}$ in the extended coordinator queue
table with the corresponding information in $S_{k}(t)$, while $M_{k}$ is the
next MP of CAV $j$.
\end{itemize}

Note that CAV $j$ may appear in multiple sub-coordinator queue tables with
different $i_{p}$ and $i_{m}$ values. However, only the one in $S_{k}(t)$
where $M_{k}$ is the next MP CAV $j$ will pass is used to update the
extended coordinator queue table $S(t)$. The information of CAV $j$ in other
sub-coordinator queue tables is necessary for determining the safety
constraints as CAV $j$ may become CAV $i_{p}$ or $i_{m}$ of other CAVs.

\textbf{Resequencing rule}: The sub-coordinator queue table allows
resequencing when a CAV passes a MP. A resequencing rule generally designs
and calculates a criterion for each CAV and sorts the CAVs in the queue
table when a new event happens. For example, FIFO takes the arrival time in
the CZ as the criterion while the Dynamic Resequencing (DR) policy\cite%
{zhang2018decentralized} uses the overall objective value in (\ref{equ:obj})
as the criterion.

We propose here a straightforward yet effective (see Section \ref{sec:simu})
resequencing rule for the roundabout as follows. Let $\tilde{x}%
_{i}^{k}\equiv x_{i}-d_{j}^{k}$ be the position of CAV $i$ relative to $M_{k}
$, where $d_{j}^{k}$ denotes the fixed distance from the entry point
(origin) $O_{j}$ to merging point $M_{k}$ along the trajectory of CAV $i$.
Then, consider 
\begin{equation}
y_{i}(t)=-\tilde{x}_{i}^{k}(t)-\varphi v_{i}(t)
\end{equation}%
This resequencing criterion reflects the distance between the CAV and the
next MP. The CAV which has the smallest $y_{i}(t)$ value is allocated first,
thus referring to this as the Shortest Distance First (SDF) policy. Note
that $\varphi v_{i}(t)$ introduces a speed-dependent term corresponding to
the speed-dependent safety constraints. This simple resequencing rule is
tested in Section \ref{sec:simu}. Other resequencing policies can also be
easily implemented with the help of the sub-coordinator queue tables.

\section{Joint optimal control and control barrier function controller (OCBF)%
}

We now return to the solution of \textbf{Problem 1}, i.e., the minimization
of \eqref{equ:obj} subject to constraints \eqref{equ:dynamics}, %
\eqref{equ:constraint1}, \eqref{equ:constraint2}, \eqref{equ:constraint3},
the initial condition $x_{i}(t_{i}^{0})=0$, and given $t_{i}^{0}$, $v_{i}^{0}
$ and $x_{i}(t_{i}^{f})$. The problem formulation is complete since we have
used the extended coordinator table to determine $i_{p}$ and $i_{m}$ (needed
for the safety cosntraints) associated with the closest MP to CAV $i$ given
the sequence of CAVs in the system. After introducing the sub-coordinator
queue tables, we also allow some resequencing for CAVs passing each MP and
focus on the CZ associated with that MP. Thus, each such problem resembles
the merging control problem in \cite{xiao2021decentralized} which can be
analytically solved. However, as pointed out in \cite{xiao2021decentralized}%
, when one or more constraints become active, this solution becomes
computationally intensive. The problem here is exacerbated by the fact that
the values of $i_{p}$ and $i_{m}$ change due to different events in the
roundabout system. Therefore, to ensure that a solution can be obtained in
real time while also guaranteeing that all safety constraints are always
satisfied, we adopt the OCBF approach which is obtained as follows: $(i)$ an
optimal control solution is first obtained for the \emph{unconstrained}
roundabout problem (as reported in \cite{xiao2021decentralized} such
solutions are computationally efficient obtain, typically requiring 
$\ll 1$sec using MATLAB). $(ii)$ This solution is used as a reference
control which is optimally tracked subject to a set of CBFs, one for each of
the constraints \eqref{equ:constraint1}, \eqref{equ:constraint2}, %
\eqref{equ:constraint3}. Using the forward invariance property of CBFs, this
ensures that these constraints are always satisfied. This whole process is
carried out in a decentralized way.

\textbf{Unconstrained optimal control solution}: With all constraints
inactive (including at $t_{i}^{0}$), the solution of \textbf{Problem 1} has
the same form as the unconstrained optimal control solution in the merging
problem \cite{xiao2021decentralized} so tha the optimal control, velocity
and position trajectories of CAV $i$ have the form: 
\begin{align}
u_{i}^{\ast }(t)& =a_{i}t+b_{i}  \label{equ:u} \\
v_{i}^{\ast }(t)& =\frac{1}{2}a_{i}t^{2}+b_{i}t+c_{i}  \label{equ:v} \\
x_{i}^{\ast }(t)& =\frac{1}{6}a_{i}t^{3}+\frac{1}{2}b_{i}t^{2}+c_{i}t+d_{i}
\label{equ:x}
\end{align}%
where the parameters $a_{i}$, $b_{i}$, $c_{i}$, $d_{i}$ and $t_{i}^{f}$ are
obtained by solving a set of nonlinear algebraic equations: 
\begin{equation}
\begin{split}
& \frac{1}{2}a_{i}\cdot (t_{i}^{0})^{2}+b_{i}\cdot t_{i}^{0}+c_{i}=v_{i}^{0},
\\
& \frac{1}{6}a_{i}\cdot (t_{i}^{0})^{3}+\frac{1}{2}b_{i}\cdot
(t_{i}^{0})^{2}+c_{i}t_{i}^{0}+d_{i}=0, \\
& \frac{1}{6}a_{i}\cdot (t_{i}^{f})^{3}+\frac{1}{2}b_{i}\cdot
(t_{i}^{f})^{2}+c_{i}t_{i}^{f}+d_{i}=x_{i}(t_{i}^{f}), \\
& a_{i}t_{i}^{f}+b_{i}=0, \\
& \beta +\frac{1}{2}a_{i}^{2}\cdot
(t_{i}^{f})^{2}+a_{i}b_{i}t_{i}^{f}+a_{i}c_{i}=0.
\end{split}%
\end{equation}

This set of equations only needs to be solved when CAV $i$ enters the CZ
and, as already mentioned, it can be done very efficiently.

\textbf{Optimal tracking controller with CBFs}: Once we obtain the
unconstrained optimal control solutions \eqref{equ:u}-\eqref{equ:x}, we use
a function $h(u_{i}^{\ast }(t),x_{i}^{\ast }(t),)x_{i}(t))$ as a control
reference $u_{ref}(t)=h(u_{i}^{\ast }(t),x_{i}^{\ast }(t),)x_{i}(t))$, where 
$x_{i}(t)$ provides feedback from the actual observed CAV trajectory). We
then design a controller that minimizes $\int_{t_{i}^{0}}^{t_{i}^{f}}\frac{1%
}{2}(u_{i}(t)-u_{ref(t)})dt$ subject to all constraints %
\eqref{equ:constraint1}, \eqref{equ:constraint2} and \eqref{equ:constraint3}%
. This is accomplished as reviewed next (see also \cite{xiao2020bridging,
xiao2019decentralized}).

First, let $\bm x_{i}(t)\equiv (x_{i}(t),v_{i}(t))$. Due to the vehicle
dynamics \eqref{equ:dynamics}, define $f(\bm x_{i}(t))=[v_{i}(t),0]^{T}$ and 
$g(\bm x_{i}(t))=[0,1]^{T}$. The constraints \eqref{equ:constraint1}, %
\eqref{equ:constraint2} and \eqref{equ:constraint3} are expressed in the
form $b_{k}(\bm x_{i}(t))\geq 0,k\in \{1,...,b\}$ where $b$ is the number of
constraints. The CBF method maps the constraint $b_{k}(\bm x_{i}(t))\geq 0$
to a new constraint which directly involves the control $u_{i}(t)$ and takes
the form 
\begin{equation}
L_{f}b_{k}(\bm x_{i}(t))+L_{g}b_{k}(\bm x_{i}(t))u_{i}(t)+\gamma (b_{k}(\bm %
x_{i}(t)))\geq 0,  \label{equ:cbf}
\end{equation}%
where $L_{f},L_{g}$ denote the Lie derivatives of $b_{k}(\bm x_{i}(t))$
along $f$ and $g$ respectively, $\gamma (\cdot )$ denotes a class of $%
\mathcal{K}$ functions \cite{xiao2019decentralized2}. The forward invariance
property of CBFs guarantees that a control input that keeps \eqref{equ:cbf}
satisfied will also keep $b_{k}(\bm x_{i}(t))\geq 0$. In other words, the
constraints \eqref{equ:constraint1}, \eqref{equ:constraint2} and %
\eqref{equ:constraint3} are never violated.

To optimally track the reference speed trajectory, a CLF function $V(\bm %
x_{i}(t))$ is used. Letting $V(\bm x_{i}(t))=(v_{i}(t)-v_{ref}(t))^{2}$, the
CLF constraint takes the form 
\begin{equation}
L_{f}V(\bm x_{i}(t))+L_{g}V(\bm x_{i}(t))u_{i}(t)+\epsilon V(x_{i}(t))\leq
e_{i}(t),  \label{equ:clf}
\end{equation}%
where $\epsilon >0$, and $e_{i}(t)$ is a relaxation variable which makes the
constraint soft. Then, the OCBF controller optimally tracks the reference
trajectory by solving the optimization problem: 
\begin{equation}
\min_{u_{i}(t),e_{i}(t)}\int_{t_{i}^{0}}^{t_{i}^{f}}\left( \beta
e_{i}^{2}(t)+\frac{1}{2}(u_{i}(t)-u_{ref(t)})\right)   \label{equ:ocbf}
\end{equation}%
subject to the vehicle dynamics \eqref{equ:dynamics}, the CBF constraints %
\eqref{equ:cbf} and the CLF constraints \eqref{equ:clf}. There are several
possible choices for $u_{ref}(t)$ and $v_{ref}(t)$. In the sequel, we choose
the simplest and most straightforward ones: 
\begin{equation}
v_{ref}(t)=v_{i}^{\ast }(t),\text{ \ \ }u_{ref}(t)=u_{i}^{\ast }(t)
\end{equation}%
where $v_{i}^{\ast }(t)$ and $u_{i}^{\ast }(t)$ are obtained from %
\eqref{equ:v} and \eqref{equ:u}.

Considering all constraints in \textbf{Problem 1}, the rear-end safety
constraint \eqref{equ:constraint1} and the vehicle limitations %
\eqref{equ:constraint3} are straightforward to transform into a CBF form. As
an example, consider \eqref{equ:constraint1} by setting $b_{1}(\bm %
x_{i}(t))=z_{i,i_{p}}(t)-\varphi v_{i}(t)-\delta $. After calculating the
Lie derivatives, the CBF constraint \eqref{equ:cbf} can be obtained. The
safe merging constraint \eqref{equ:constraint2} differs from the rest in
that it only applies to specific time instants $t_{i}^{m_{k}}$. This poses a
technical complication due to the fact that a CBF must always be in a
continuously differential form. We can convert \eqref{equ:constraint2} to
such a form using the technique in \cite{xiao2019decentralized2} to obtain 
\begin{equation}
z_{i,i_{m}}(t)-\Phi (x_{i}(t))v_{i}(t)-\delta \geq 0,\text{ \ }t\in \lbrack
t_{i}^{k,0},t_{i}^{k}]  \label{equ:cbf2}
\end{equation}%
where $t_{i}^{m_{k,0}}$ denotes the time CAV $i$ enters the road segment
connected to $M_{k}$ and $\Phi (\cdot )$ is any strictly increasing function
as long as it satisfies the boundary constraints $z_{i,i_{m}}(t_{i}^{k,0})-%
\phi v_{i}(t_{i}^{k,0})-\delta \geq 0$ and $z_{i,i_{m}}(t_{i}^{k})-\phi
v_{i}(t_{i}^{k})-\delta \geq 0$ (which is precisely \eqref{equ:constraint2}%
). Note that we need to satisfy \eqref{equ:cbf2} when a CAV changes road
segments in the roundabout and the value of $i_{m}$ changes. Since $%
z_{i,i_{m}}(t_{i}^{k,0})\geq -L_{i_{m}}+L_{i}$, where $L_{i}$ is the length
of the road segment CAV $i$ is in, to guarantee the feasibility of %
\eqref{equ:cbf2}, we set $\Phi (x_{i}(t_{i}^{k,0}))v_{i}(t_{i}^{k,0})+\delta
=-L_{i_{m}}+L_{i}$. Then, from \eqref{equ:constraint2}, we get $\Phi
(x_{i}(t_{i}^{k}))=\varphi $. Simply choosing a linear $\Phi (\cdot )$ as
follows: 
\begin{equation}
\Phi (x_{i}(t))=\left( \varphi +\frac{L_{i_{m}}-L_{i}+\delta }{%
v_{i}(t_{i}^{k,0})}\right) \frac{x_{i}(t)}{L_{i}}-\frac{L_{i_{m}}-L_{i}+%
\delta }{v_{i}(t_{i}^{k,0})}
\end{equation}%
it is easy to check that it satisfies the boundary requirements. Note that
when implementating the OCBF controller, $x_{i}(t)$ needs to be transformed
into a relative position $\tilde{x}_{i}^{k}+L_{i}$, which reflects the
distance between CAV $i$ and the origin of the current road segment. Then, $%
z_{i,i_{p}}$ and $z_{i,i_{m}}$ are calculated after this transformation,
where $z_{i,i_{p}}=\tilde{x}_{i_{p}}^{k}-\tilde{x}_{i}^{k}$, $z_{i,i_{p}}=%
\tilde{x}_{i_{m}}^{k}-\tilde{x}_{i}^{k}$.

With all constraints converted to CBF constraints in \eqref{equ:ocbf}, we
can solve this problem by discretizing $[t_{i}^{0},t_{i}^{f}]$ into
intervals of equal length $\Delta $ and solving \eqref{equ:ocbf} over each
time interval. The decision variables $u_{i}(t)$ and $e_{i}(t)$ are assumed
to be constant on each such time interval and can be easily obtained by
solving a Quadratic Program (QP) problem since all CBF constraints are
linear in the decision variables $u_{i}(t)$ and $e_{i}(t)$. By repeating
this process until CAV $i$ exit the CZ, the solution to \eqref{equ:ocbf} is
obtained.

Algorithm \ref{alg:1} summarizes the overall process for solving the
roundabout problem: each CAV $i$ first determines the conflict CAVs $i_{p}$
and $i_{m}$ and then derives the OCBF controller that guides it through its
trajectory.

\begin{algorithm}[htb]
	\SetAlgoLined
	\For{every T seconds}{
		\If{a new event occurs}{
			\If{new CAV entering event}{
				Determine the passing order for all CAVs\\
				Plan an unconstrained optimal control trajectory for the new CAV
			}
			Update the extended coordinator queue table $S(t)$
		}
		\For{each CAV in S(t)}{
			Determine the safety contraints it needs to meet \\
			Use the joint optimal control and barrier function controller to obtain control for it
		}
	}
	\caption{A MP-based Algorithm for Roundabout Problems}
	\label{alg:1}
\end{algorithm}

\section{Simulation Results}

\label{sec:simu} In this section, we use Vissim, a multi-model traffic flow
simulation platform, as a baseline to compare a roundabout perfpormance with
humna-driven vehicles to our OCBF controller. We build the scenario shown in
Fig. \ref{fig:model} in Vissim and use the same vehicle arrival patterns in
the OCBF controller for consistent comparison purposes.

\textbf{Simulation 1}: The first simulation focuses on the performance of
the OCBF controller. The basic parameter settings are as follows: $L_{a}=60m$%
, $L=60m$, $\delta =10m$, $\varphi =1.8s$, $v_{\max }=17m/s$, $v_{\min }=0$, 
$u_{\max }=5m/s^{2}$, $u_{\min }=-5m/s^{2}$. This scenario considers a \emph{%
symmetric} configuration in the sense that $L_{a}=L$. The traffic in the
three incoming roads is generated through Poisson processes with all rates
set to $360$ CAVs/h. Under these traffic rates, vehicles will sometimes line
up waiting for other vehicles in the roundabout to pass. A total number of
approximately 200 CAVs are simulated. The simulation results of the
performance of OCBF compared to that in Vissim are listed in Table \ref%
{tab:result}.

\begin{table}[htb]
\centering
\begin{threeparttable}
		\caption{Objective function comparison for a symmetric roundabout}
		\label{tab:result}
		\begin{tabular}{ccccc}
			\toprule
			Items & \multicolumn{2}{c}{OCBF} & \multicolumn{2}{c}{Vissim} \\
			\midrule
			Weight & $\alpha = 0.1$ & $\alpha = 0.2$ & $\alpha = 0.1$ & $\alpha = 0.2$ \\
			\midrule
			Ave. time (s) & 13.7067 & 13.3816 & \multicolumn{2}{c}{20.6772} \\
			Ave. energy & 16.0698 & 24.6336 & \multicolumn{2}{c}{33.2687} \\
			Ave. obj.\tnote{1} & 35.1084 & 66.4511 & 61.9893 & 97.8850 \\
			\bottomrule
		\end{tabular}
		\begin{tablenotes}
			\item [1] Ave. obj = $\beta \times$ Ave. time + Ave. energy, \\$\beta = \frac{\alpha\max\{u_{\max{}}^2, u_{\min{}}^2\}}{2(1-\alpha)}$
		\end{tablenotes}
	\end{threeparttable}
\end{table}

In this simulation, FIFO is chosen as the sequencing policy in the OCBF
method. As seen in Table \ref{tab:result}, the travel time of CAVs in the
roundabout improves about 34\% using the OCBF method compared with that of
Vissim when $\alpha =0.1$ (with some additional improvement when $\alpha =0.2
$). The CAVs using the OCBF method consume 52\% and 26\% less energy than
that in Vissim with $\alpha $ set to 0.1 and 0.2 respectively. A larger $%
\alpha $ means more emphasis on the travel time than the energy consumption,
which explains the shorter travel time and the larger energy consumption.
When it comes to the total objective, the OCBF controller shows 44\% and
32\% improvement over the human-driven perfprmance in Vissim when $\alpha $
equals to 0.1 and 0.2 respectively. This improvement in both the travel time
and the energy consumption is to be expected as the CAVs using the OCBF
method never stop and wait for CAVs in another road segment to pass as
Vissim do.

\textbf{Simulation 2}: The second simulation compares the performance of
OCBF under different sequencing rules in an \emph{asymmetric} configuration.
The parameter settings are the same as the first case except that $L=100m$.
The weight is set to $\alpha =0.2$. The simulation results of the
performance of OCBF with FIFO and OCBF with the SDF sequencing policy, as
well that in Vissim, are shown in Table. \ref{tab:result2}.

\begin{table}[tbh]
\caption{Objective function comparison of different resequencing rule for an
asymmetric roundabout}
\label{tab:result2}\centering
\begin{tabular}{cccc}
\toprule Items & OCBF+FIFO & OCBF+SDF & Vissim \\ 
\midrule Ave. time (s) & 16.4254 & 14.7927 & 24.6429 \\ 
Ave. energy & 56.9643 & 23.1131 & 30.8947 \\ 
Ave. obj. & 108.2937 & 69.3403 & 107.9038 \\ 
\bottomrule &  &  & 
\end{tabular}
\end{table}

Table \ref{tab:result2} shows that a CAV using OCBF with FIFO spends around
33\% less travel time but 84\% more energy than that in Vissim. The average
objective values of the two cases are almost the same, indicating that OCBF
with FIFO works poorly in an asymmetric roundabout. For example, when a CAV
enters segment $l_{4}$, it has to wait for another CAV that has entered $%
l_{2}$ just before it to run 40 more meters for safe merging. This is
unreasonable and may also result in some extreme cases when the OCBF problem
becomes infeasible. This problem can be resolved by choosing a better
sequencing policy such as SDF. As shown in Table \ref{tab:result2}, OCBF+SDF
outperforms OCBF+FIFO, achieving an improvement of 40\% in travel time, 26\%
in energy consumption and 36\% in the objective value compared to that in
Vissim.

\textbf{Simulation 3}: The purpose of this experiment is to study the effect
of traffic volume. The roundabout is set to be \emph{asymmetric} with the
same parameter settings as in \textbf{Simulation 2}. A total number of
approximately 500 CAVs are simulated under two sets of incoming traffic
rates: balanced incoming traffic (360 CAVs/h for each origin) and imbalanced
incoming traffic (540 CAVs/h from $O_{1}$, 270 CAVs/h from $O_{2}$ and $O_{3}
$). The simulation results of the performance of OCBF+SDF compared to that
in Vissim under both balanced and imbalanced incoming traffic are shown in
Tables \ref{tab:result3} and \ref{tab:result4}. 
\begin{table}[tbh]
\caption{Objective function comparison (Vissim)}
\label{tab:result3}\centering
\begin{tabular}{ccccc}
\toprule Traffic Type & CAV Origin & Time & Energy & Ave. Obj. \\ 
\midrule \multirow{4}{*}{Balanced} & All & 24.7615 & 32.1295 & 109.5092 \\ 
& From $O_1$ & 24.7631 & 33.7724 & 111.1570 \\ 
& From $O_2$ & 23.6934 & 31.4432 & 105.4851 \\ 
& From $O_3$ & 25.8458 & 31.0375 & 111.8057 \\ 
\midrule \multirow{4}{*}{Imbalanced} & Total & 26.5170 & 36.0547 & 118.9203
\\ 
& From $O_1$ & 28.1765 & 39.5067 & 127.5582 \\ 
& From $O_2$ & 25.6977 & 35.5803 & 115.8857 \\ 
& From $O_3$ & 23.8207 & 29.2346 & 103.6744 \\ 
\bottomrule &  &  &  & 
\end{tabular}%
\end{table}
\begin{table}[tbh]
\caption{Objective function comparison (OCBF+SDF)}
\label{tab:result4}\centering
\begin{tabular}{ccccc}
\toprule Traffic Type & CAV Origin & Time & Energy & Ave. Obj. \\ 
\midrule \multirow{4}{*}{Balanced} & All & 15.1650 & 24.8601 & 72.2507 \\ 
& From $O_1$ & 15.1613 & 24.8453 & 72.2245 \\ 
& From $O_2$ & 14.8221 & 25.3610 & 71.6801 \\ 
& From $O_3$ & 15.5156 & 24.3696 & 72.8560 \\ 
\midrule \multirow{4}{*}{Imbalanced} & Total & 15.2437 & 24.1693 & 71.8059
\\ 
& From $O_1$ & 15.4035 & 24.9519 & 73.0880 \\ 
& From $O_2$ & 14.6235 & 19.8901 & 65.5885 \\ 
& From $O_3$ & 15.5163 & 26.7188 & 75.2072 \\ 
\bottomrule &  &  &  & 
\end{tabular}%
\end{table}

From Table \ref{tab:result3}, it is seen that the imbalanced traffic causes
longer travel times ($\sim $2s) and more energy consumption ($\sim $13\%)
although the total traffic rates are the same. The imbalanced traffic
results in an imbalanced performance of CAVs from different origins. The
CAVs originated from $O_{1}$ with heavy traffic perform worse than those
from $O_{2}$ and $O_{3}$ with light traffic. However, when OCBF+SDF is
applied to the system, the imbalanced traffic brings almost no performance
loss and becomes somewhat more balanced after passing the roundabout. This
result is interesting because the fact that traffic is \emph{imbalanced} was
not taken into consideration in our method. An explanation of this
phenomenon is that SDF gives the CAVs from $O_{1}$ a higher priority as they
are more likely to be the closest to the MP while OCBF allows the CAVs to
pass the roundabout quickly without stop; therefore, the CAVs from a heavy
traffic flow are less likely to get congested.

\section{Conclusion}

We have presented a decentralized optimal control framework for controlling
CAVs traveling through a roundabout to jointly minimize both the travel time
and the energy consumption while satisfying speed-dependent safety
constraints, as well as velocity and acceleration constraints. A systematic
method is designed to dynamically determine the safety constraints a CAV
needs to meet. An OCBF controller, combining an unconstrained optimal
control solution with CBFs, is designed and implemented to track the desired
(unconstrained) trajectory while guaranteeing that all safety constraints
and vehicle limitations are satisfied. Significant improvements are shown in
the simulation experiments which compare the performance of the OCBF
controller to a baseline of human-driven vehicles. Future research is
directed at studying different sequencing policies (which our resuls show is
important in asymmetric roundabout configurations), as well as considering
the centrifugal discomfort caused when road segments are curved and
extending the model to more complex roudabouts as well as to a multi-lane
version which allows lane changing and overtaking.

\bibliographystyle{ieeetr}
\bibliography{cav}

\begin{thebibliography}{10}

\bibitem{rios2016survey}
J.~Rios-Torres and A.~A. Malikopoulos, ``A survey on the coordination of
  connected and automated vehicles at intersections and merging at highway
  on-ramps,'' {\em IEEE Transactions on Intelligent Transportation Systems},
  vol.~18, no.~5, pp.~1066--1077, 2016.

\bibitem{chen2015cooperative}
L.~Chen and C.~Englund, ``Cooperative intersection management: A survey,'' {\em
  IEEE Transactions on Intelligent Transportation Systems}, vol.~17, no.~2,
  pp.~570--586, 2015.

\bibitem{tideman2007review}
M.~Tideman, M.~C. van~der Voort, B.~van Arem, and F.~Tillema, ``A review of
  lateral driver support systems,'' in {\em 2007 IEEE Intelligent
  Transportation Systems Conference}, pp.~992--999, IEEE, 2007.

\bibitem{li2013survey}
L.~Li, D.~Wen, and D.~Yao, ``A survey of traffic control with vehicular
  communications,'' {\em IEEE Transactions on Intelligent Transportation
  Systems}, vol.~15, no.~1, pp.~425--432, 2013.

\bibitem{xu2019grouping}
H.~Xu, S.~Feng, Y.~Zhang, and L.~Li, ``A grouping-based cooperative driving
  strategy for cavs merging problems,'' {\em IEEE Transactions on Vehicular
  Technology}, vol.~68, no.~6, pp.~6125--6136, 2019.

\bibitem{xu2020bi}
H.~Xu, Y.~Zhang, C.~G. Cassandras, L.~Li, and S.~Feng, ``A bi-level cooperative
  driving strategy allowing lane changes,'' {\em Transportation research part
  C: emerging technologies}, vol.~120, p.~102773, 2020.

\bibitem{milanes2010automated}
V.~Milan{\'e}s, J.~Godoy, J.~Villagr{\'a}, and J.~P{\'e}rez, ``Automated
  on-ramp merging system for congested traffic situations,'' {\em IEEE
  Transactions on Intelligent Transportation Systems}, vol.~12, no.~2,
  pp.~500--508, 2010.

\bibitem{rios2015online}
J.~Rios-Torres, A.~Malikopoulos, and P.~Pisu, ``Online optimal control of
  connected vehicles for efficient traffic flow at merging roads,'' in {\em
  2015 IEEE 18th international conference on intelligent transportation
  systems}, pp.~2432--2437, IEEE, 2015.

\bibitem{bichiou2018developing}
Y.~Bichiou and H.~A. Rakha, ``Developing an optimal intersection control system
  for automated connected vehicles,'' {\em IEEE Transactions on Intelligent
  Transportation Systems}, vol.~20, no.~5, pp.~1908--1916, 2018.

\bibitem{hult2016coordination}
R.~Hult, G.~R. Campos, E.~Steinmetz, L.~Hammarstrand, P.~Falcone, and
  H.~Wymeersch, ``Coordination of cooperative autonomous vehicles: Toward safer
  and more efficient road transportation,'' {\em IEEE Signal Processing
  Magazine}, vol.~33, no.~6, pp.~74--84, 2016.

\bibitem{cao2015cooperative}
W.~Cao, M.~Mukai, T.~Kawabe, H.~Nishira, and N.~Fujiki, ``Cooperative vehicle
  path generation during merging using model predictive control with real-time
  optimization,'' {\em Control Engineering Practice}, vol.~34, pp.~98--105,
  2015.

\bibitem{mukai2017model}
M.~Mukai, H.~Natori, and M.~Fujita, ``Model predictive control with a mixed
  integer programming for merging path generation on motor way,'' in {\em 2017
  IEEE Conference on Control Technology and Applications (CCTA)},
  pp.~2214--2219, IEEE, 2017.

\bibitem{nor2018merging}
M.~H. B.~M. Nor and T.~Namerikawa, ``Merging of connected and automated
  vehicles at roundabout using model predictive control,'' in {\em 2018 57th
  Annual Conference of the Society of Instrument and Control Engineers of Japan
  (SICE)}, pp.~272--277, IEEE, 2018.

\bibitem{xiao2020bridging}
W.~Xiao, C.~G. Cassandras, and C.~Belta, ``Bridging the gap between optimal
  trajectory planning and safety-critical control with applications to
  autonomous vehicles,'' {\em Automatica}, 2021 (in print).

\bibitem{xiao2019decentralized}
W.~Xiao, C.~G. Cassandras, and C.~Belta, ``Decentralized merging control in
  traffic networks with noisy vehicle dynamics: A joint optimal control and
  barrier function approach,'' in {\em 2019 IEEE Intelligent Transportation
  Systems Conference (ITSC)}, pp.~3162--3167, IEEE, 2019.

\bibitem{zhang2019decentralized}
Y.~Zhang and C.~G. Cassandras, ``Decentralized optimal control of connected
  automated vehicles at signal-free intersections including comfort-constrained
  turns and safety guarantees,'' {\em Automatica}, vol.~109, p.~108563, 2019.

\bibitem{xiao2021decentralized}
W.~Xiao and C.~G. Cassandras, ``Decentralized optimal merging control for
  connected and automated vehicles with safety constraint guarantees,'' {\em
  Automatica}, vol.~123, p.~109333, 2021.

\bibitem{flannery1997operational}
A.~Flannery and T.~Datta, ``Operational performance measures of american
  roundabouts,'' {\em Transportation research record}, vol.~1572, no.~1,
  pp.~68--75, 1997.

\bibitem{martin2016benefits}
M.~Martin-Gasulla, A.~Garc{\'\i}a, and A.~T. Moreno, ``Benefits of metering
  signals at roundabouts with unbalanced flow: Patterns in spain,'' {\em
  Transportation Research Record}, vol.~2585, no.~1, pp.~20--28, 2016.

\bibitem{yang2004new}
X.~Yang, X.~Li, and K.~Xue, ``A new traffic-signal control for modern
  roundabouts: method and application,'' {\em IEEE Transactions on Intelligent
  Transportation Systems}, vol.~5, no.~4, pp.~282--287, 2004.

\bibitem{xu2016multi}
H.~Xu, K.~Zhang, and D.~Zhang, ``Multi-level traffic control at large four-leg
  roundabouts,'' {\em Journal of Advanced Transportation}, vol.~50, no.~6,
  pp.~988--1007, 2016.

\bibitem{zhao2018optimal}
L.~Zhao, A.~Malikopoulos, and J.~Rios-Torres, ``Optimal control of connected
  and automated vehicles at roundabouts: An investigation in a mixed-traffic
  environment,'' {\em IFAC-PapersOnLine}, vol.~51, no.~9, pp.~73--78, 2018.

\bibitem{chalaki2020experimental}
B.~Chalaki, L.~E. Beaver, and A.~A. Malikopoulos, ``Experimental validation of
  a real-time optimal controller for coordination of cavs in a multi-lane
  roundabout,'' in {\em 2020 IEEE Intelligent Vehicles Symposium (IV)},
  pp.~775--780, IEEE, 2020.

\bibitem{xiao2020decentralized}
W.~{Xiao}, C.~G. {Cassandras}, and C.~{Belta}, ``Decentralized optimal control
  in multi-lane merging for connected and automated vehicles,'' in {\em 2020
  IEEE 23rd International Conference on Intelligent Transportation Systems
  (ITSC)}, pp.~1--6, 2020.

\bibitem{xu2020comparison}
H.~Xu, C.~G. Cassandras, L.~Li, and Y.~Zhang, ``Comparison of cooperative
  driving strategies for cavs at signal-free intersections,'' {\em submitted to
  IEEE Transactions on Intelligent Transportation Systems}, 2020.

\bibitem{zhang2018decentralized}
Y.~Zhang and C.~G. Cassandras, ``A decentralized optimal control framework for
  connected automated vehicles at urban intersections with dynamic
  resequencing,'' in {\em 2018 IEEE Conference on Decision and Control (CDC)},
  pp.~217--222, IEEE, 2018.

\bibitem{vogel2003comparison}
K.~Vogel, ``A comparison of headway and time to collision as safety
  indicators,'' {\em Accident analysis \& prevention}, vol.~35, no.~3,
  pp.~427--433, 2003.

\bibitem{xiao2019decentralized2}
W.~Xiao, C.~Belta, and C.~G. Cassandras, ``Decentralized merging control in
  traffic networks: A control barrier function approach,'' in {\em Proceedings
  of the 10th ACM/IEEE International Conference on Cyber-Physical Systems},
  pp.~270--279, 2019.

\end{thebibliography}
{}

\end{document}